A versatile electrophoresis system for the analysis of high and low molecular weight proteins

Christophe Tastet[1], Pierre Lescuyer[1], Hélène Diemer[2], Sylvie Luche[1], Alain van Dorsselaer[2] and Thierry Rabilloud[1]

1: CEA- Laboratoire de Bioénergétique Cellulaire et Pathologique, EA 2943, DRDC/BECP CEA-Grenoble, 17 rue des martyrs, F-38054 GRENOBLE CEDEX 9, France

2: Laboratoire de Spectrométrie de Masse Bio-Organique, UMR CNRS 7509, ECPM, 2( rue Becquerel, 67008 Strasbourg Cedex

Correspondence :
Thierry Rabilloud, DRDC/BECP
CEA-Grenoble, 17 rue des martyrs,
F-38054 GRENOBLE CEDEX 9
Tel (33)-4-38-78-32-12
Fax (33)-4-38-78-51-87
e-mail: Thierry.Rabilloud@ cea.fr

Abbreviations:
AMPSO: (1,1 dimethyl-2-hydroxy ethyl)-3-aminp-2-hydroxy-propane sulfonic acid

Abstract

A new, versatile, multiphasic buffer system for high resolution sodium dodecyl sulfate-polyacrylamide gel electrophoresis of proteins in the relative molecular weight Mw range of 300,000-3000 Da is described. The system, based on the theory of multiphasic zone electrophoresis, allows complete stacking and destacking of proteins in the above Mw range. The buffer system uses taurine and chloride as trailing and leading ion, respectively, and Tris, at a pH close to its pKa, as the buffering counter ion. Coupled with limited variation in the acrylamide concentration, this electrophoresis system allows to tailor the resolution in the 6-200 kDa Mw range, with minimal difficulties in the post electrophoretic identification processes.

Introduction

In almost all 2D electrophoresis systems, the SDS PAGE dimension uses the classical Tris-chloride-glycine discontinuous system introduced by Ornstein and Davis [1]. While this system gives excellent performances for proteins above 15-20 kDa, its performances decreases for proteins of lower molecular weight. To be analyzed with this electrophoresis system, high concentration gels are needed (example in [2]) which poses extra problems. These high concentration gels are very brittle and thus difficult to handle. In addition, as the pore size of the polyacrylamide gel is very small, penetration of reactants in the gel after electrophoresis is more difficult. This is especially true when high molecular weight reactants such as proteolytic enzymes are used, as it is the case in the peptide mass fingerprinting approaches. This example points out the limits of playing with the acrylamide concentration for modulating the electrophoresis system. Fortunately, discontinuous electrophoresis offers another possibility for modulating the separation power, which is the buffer system. According to the basic theory of discontinuous electrophoresis [3], the speed of the moving boundary depends on a complex relationship between the pH of the gel, the pKas of the leading and trailing ions and the intrinsic mobility of the leading and trailing ions. Most generally, strong acids counterions (most commonly chloride) are used as leading ions, and the mobility of the moving boundary is not heavily modulated by changing those ions. Thus, the main players are the pH of the gel and the pKa and mobility of the trailing ion. As a general rule, with a given trailing ion, the mobility of the moving boundary will increase with increasing pH. This will allow in turn unstacking of more mobile species (i.e. low molecular weight proteins in SDS systems) from the boundary, and thus allow resolution of low molecular weight proteins from the buffer front at constant acrylamide concentration. Conversely, lowering the pH of the gel will slow down the speed of the mowing boundary, keeping therefore low molecular weight protein in the buffer front and leaving more gel space for the analysis of higher molecular weight proteins, with increased resolution in the corresponding Mw ranges [4]. In the case of the glycine-based system, only pH lowering is practical with a Tris buffer, as the standard pH in the resolving gel is 8.8, which is already rather far from the 8.05 pKa of Tris. Thus, buffers of higher pK have been proposed for the electrophoresis of low molecular weight proteins [5]. However, this approach is not trouble-free, as the real running pH begins to be above 9.5, with induced hydrolysis of the acrylamide and increased electroendosmosis.

Thus, the most common approach in this case is the other modulating factor, i.e. the nature of the trailing ion. Uusally, trailing ions with a carboxylic acid function (generally glycine derivatives) and a basic function with a lower pKa are used. However, trailing ions with sulfonic acidic groups have also been used [6]. Such trailing ions allow high moving boundary speeds at reasonable pH values, and are therefore very convenient for the analysis of low molecular weight proteins and peptides [7] [8]. However, their use can induce some problems. For example, some silver staining techniques (e.g. ammoniacal silver) are not compatible with Tricine and/or Bicine gels [9]. In addition, these systems are not suited for the high-resolution analysis of medium and high molecular weight proteins. As a matter of facts, they require low pH values (e.g. close to 7 for Tricine) to give the adequate boundary speed to mimic the performances of Tris-glycine system [10]. These relatively low pH values are in turn almost outside the buffering pH range of Tris, thereby requiring different buffering species (e.g. BisTris) which are more expensive and quite often interfere either with acrylamide polymerization and/or with some silver staining procedures. This also means that two different buffers systems must be used for the analysis of high and low molecular weight proteins. We therefore looked for a more versatile system which would allow for wide modulation of the moving boundary speed with a single buffering ion and would also be fully compatible with the major types of silver staining.

Materials and methods

Cell culture and oxidative stress

Hela S3 cells were cultured in suspension in DMEM medium containing 1mM pyruvate, 10mM Hepes-NaOH pH 7.5 and 10% fetal calf serum. Cells were then harvested by centrifugation, rinsed in PBS and resuspended in homogeneization buffer (0.25 M sucrose, 10 mM Tris-HCl, pH 7.5, 1 mM EDTA). A buffer volume approximately equal to the packed cell volume was used. The suspension was transferred to a polyallomer ultracentrifuge tube and the cells were lysed by the addition of 4 volumes (respective to the suspension volume) of 8.75 M urea, 2.5 M thiourea, 25 mM spermine base and 50 mM DTT. After 1 hour at room temperature, the extracts were ultracentrifuged (30 minutes at 200000g). The supernatant was collected and the protein was determined by a Bradford assay, using bovine serum albumin as a standard. Carrier ampholytes (0.4 % final concentration) were added and the protein extracts were stored at -20°C.

Two-dimensional electrophoresis

Two-dimensional electrophoresis was performed with immobilised pH gradients for isoelectric focusing. Home-made linear 4-8 or 4-12 gradients were used [11], and prepared according to published procedures [12]. IPG strips were cut with a paper cutter, and rehydrated in 7M urea, 2M thiourea, 4% CHAPS, 0.4% carrier ampholytes (3-10 range), containing either 20mM DTT (4-8 gradients) or 5mM Tris cyanoethyl phosphine (purchased from MolecularProbes, for 4-12 gradients) [13]. The protein sample was mixed with the rehydration solution in the case of 4-8 gradients, or cup-loaded at the anode for 4-12 gradients. Isoelectric focusing was carried out for a total of 60000 Vh. After focusing, the strips were equilibrated for 2 x 10 minutes in 6M urea, 2% SDS, 125 mM Tris-HCl pH 7.5 containing either 50mM DTT (first equilibration step) or 150mM iodoacetamide (second equilibration step) [14] . The equilibrated strip was loaded on the top of a 10% polyacrylamide gel, and submitted to SDS PAGE at 12W/ gel.

SDS-PAGE

For SDS PAGE, various buffer systems were evaluated. As a general rule, buffers were prepared from pH calculations by weighing Tris base and measuring volumes of standard solutions of hydrochloric acid (Fluka). Thus, buffers are designated by the mass of Tris and the HCl concentration present in the concentrated buffer. As an example Tris 120/0.6 means 120 g of Tris per liter and 0.6M HCl. At such high Tris concentrations, the pH values given by most pHmeters are rather erratic, so that the theoretical pH value is given for convenience reasons. This pH value is in addition more representative of the real pH value after dilution in the final acrylamide gel. As a rule in comparing buffers of various pH, we decided to operate at constant ionic strength, corresponding to 0.1M HCl (final concentration in gel). When needed, stacking gels (4%T) were cast in Tris-HCl pH 7.5 (Tris 90/0.6).

Various trailing ions were tested as electrode buffer components. In addition to glycine, Tricine, asparagine, AMPSO and taurine were tested. Typical electrode buffers contain 0.1% SDS, 50mM Tris and 200mM of trailing ion. Various acrylamide concentrations were also tested, from 9 to 15% T (at constant 2.7%C).

Molecular weight standards were from Bio Rad (broad range unstained standards, 200-6.5 kDa range) or Sigma (17-2.5 kDa range). For high molecular mass determinations, crosslinked standards were prepared as follows. Bovine serum albumin was dissolved at 5 mg/ml in 100mM phosphate buffer, pH 7. Pyromellitic dianhydride (predissolved at 1M in dimethylformamide) was added at 20mM final concentration. Crosslinking took place at 4°C for 18 hours. the reaction was then stopped by addition of Tris buffer pH 7.5 (0.2M final concentration). The protein samples were diluted in sample buffer (Tris HCl pH 7.5, 2% SDS, 50mM dithiothreitol, 15% glycerol) and denatured by heating at 100°C for 5 minutes.

After migration, the gels were stained either with silver nitrate [15] for 2D gels with a pH 4-8 gradient, or with ammoniacal silver for 2D gels with a pH 5.5 to 12 gradient [16], or with a fluorescent ruthenium complex [17] when subsequent mass spectrometry on excised spots was performed.

Mass spectrometry

In gel digestion : Spot digestion was performed essentially as described previously [17]. Stained proteins spots or bands were excised and shrunk in 1 ml of 50% ethanol for 2 hours. Each gel slice was cut into small pieces with a scalpel, washed with 100 $\mu$l of 25 mM NH4HCO3 and dehydrated with 100 $\mu$l of acetonitrile. This operation was repeated twice. Gel pieces were completely dried with a Speed Vac before enzymatic digestion. The dried gel volume was evaluated and three volumes of trypsin (12.5 ng/$\mu$l) in 25 mM NH4HCO3 (freshly diluted) were added. The digestion was performed at 35°C overnight. The gel pieces were centrifuged and 5 $\mu$l of 25% H2O/70% Acetonitrile/5% HCOOH were added. The mixture was sonicated for 5 min. and centrifuged. The supernatant was recovered and the operation was repeated once. For MALDI-MS analysis, the supernatant volume was reduced under nitrogen flow to 4 $\mu$l, 1 $\mu$l of H2O/5% HCOOH were added and 0.5 $\mu$l of the mix were used for the analysis.

MALDI-MS : For MALDI mass spectrometry, mass measurements were carried out on a Bruker BIFLEX III™ MALDI-TOF equipped with the SCOUT™ High Resolution Optics with X-Y multisample probe and gridless reflector. This instrument was used at a maximum accelerating potential of 20 kV (in positive mode) or – 20 kV (in negative mode) and was operated in reflector mode. A saturated solution of α-cyano-4-hydroxycinnamic acid in acetone was used as a matrix. A first layer of fine matrix crystals was obtained by spreading and fast evaporation of 0.5 $\mu$l of matrix solution. On this fine layer of crystals, a droplet of 0.5 $\mu$l of aqueous HCOOH (5%) solution was deposited. Afterwards, 0.5 $\mu$l of sample solution was added and a second 0.2 $\mu$l droplet of saturated matrix solution (in 50% H2O/50% acetonitrile) was added. The preparation was dried under vacuum. The sample was washed one to three times by applying 1$\mu$l of aqueous HCOOH (5 %) solution on the target and then flushed after a few seconds. Internal calibration was performed with Angiotensin, Substance P, Bombesin, and ACTH, with respectively monoisotopic masses at m/z = 1046.542; m/z = 1347.736; m/z = 1620.807; m/z = 2465.199.

The MS-FIT program from the ProteinProspector package (University of California at San Francisco) was used to database searching (http://prospector.ucsf.edu). All proteins present in Swiss-Prot and TrEMBL were taken into account without any pI and Mr restrictions. The peptide mass error was limited to 100 ppm, 1 miss cleavage might be accepted and no AA substitutions were allowed.

Results and discussion

From our premises of a single buffer system allowing maximal versatility, additional constraints appeared almost immediately. We selected Tris as the buffering component, both for economical reasons, because it gives excellent acrylamide polymerization with excellent gel structure, and because it has proved compatible with all silver staining protocols published to date. This initial choice strongly restricted the range of trailing ions to be tested. This choice was further restricted by our decision to have a buffer system compatible with the silver nitrate techniques, which give high quality patterns for acidic and neutral proteins and with the ammoniacal silver staining techniques, which give superior results for basic proteins [18]. It soon became apparent that the latter techniques offer a very poor choice of buffering and trailing ions. Both the morpholine derivatives (e.g. MES, MOPS) and the hydroxylated compounds (e.g. Tricine, Bicine, triethanolamine), with the notable exception of Tris and BisTris, gave an intense yellow-brown background with ammoniacal silver techniques. This meant in turn that only trailing ions with a primary amino group, such as the one present in amino acids, are really compatible with ammoniacal silver staining.

We were thus restricted to a trailing ion with a primary amino group with a pKa which had to be much lower than the one of glycine (ca. 9.7) in order to afford moving boundary speed modulation around the pKa of Tris (ca. 8). Three simple ions met these requirements: AMPSO, asparagine and taurine. These three ions were tested and performed adequately (data not shown). Taurine was finally selected on economical grounds, as the amount of trailing ion used is rather high, even if its presence is mandatory only in the cathode buffer. As a matter of facts, in the systems were the cathode and anode buffer are separated, the standard and cheap Tris-glycine-SDS buffer can be used in the anode compartment of the electrophoresis setup.

Figure 1 shows the migration patterns obtained on markers proteins with the taurine-based system, compared to the standard glycine system and to the well-known tricine system. It can be easily seen that the taurine system affords great versatility with minimal action on the pH and acrylamide percentage. The third vertical column (panels C, F and I), shows the influence of the pH of the resolving gel at constant %T (10%) on the separation. The pH changed from 7.75 to 8.05 only, but important changes in the speed of the moving boundary are experienced, with corresponding changes in the Mw range separated in the gels. While the 14 kDa marker is barely separated from the buffer front at pH 7.75, it is clearly unstacked at pH 7.95. The effect is even more dramatic at pH 8.05, where the 6.5kDa marker is clearly unstacked from the buffer front in a 10% gel. As a result, the taurine system affords with a single pH change in the resolving gel either a resolution comparable to the glycine system (panel D vs F) or to the Tricine system (panel E vs I). Further modulation of the moving boundary speed was attempted by wider variation of the pH, with very limited success. As a matter of facts, the discontinuous electrophoresis theory [3] shows that the moving boundary speed vs. pH curve is not linear, with a wide variation over a limited pH range and much shallower variations outside this optimal pH range. This implies in turn that the preparation of the gel buffers must be extremely standardized to avoid erratic migration patterns. We therefore strongly advise against the pH titration practice, as the inherent pH meter imprecision at these high Tris concentrations, coupled with the effect of the heat generated by the titration of Tris base with concentrated HCl are very likely to give erratic results. This is why we selected a procedure without pH titration, using Tris base and commercial standard HCl solutions (1 or 2M). Alternatively, a procedure using Tris base and Tris hydrochloride is equally efficient.

In addition, the separation can be widely modulated by moderate changes in the acrylamide concentration. For example, a simple change from 10 to 11.5 %T allows to unstack a 2.5 kDa peptide from the moving boundary and thus allows good resolution of relatively low molecular weight peptides. This is achieved without going up too much in acrylamide concentration, which induces in turn brittle gels and problems in the moving boundary zone linked to unstacking of the SDS micelles (see panels G and H, bottom of the gels). Conversely, minimal decrease in the acrylamide concentration (from 10 to 9%) allocates more space for the resolution of high molecular weight proteins (up to 200 kDa) without going to low in acrylamide concentration, which leads to fragile gels. If the analysis is devoted completely to high molecular weight proteins, lowering the gel concentration to 7.5% T allows to analyse proteins up to 300 kDa at a reasonable gel strength.

It can be objected that the resolution scope described with the taurine system can be obtained with a combination of the glycine and tricine system. While this is true in principle, the taurine system offers many practical advantages. First of all, it is completely compatible with all types of staining, including ammoniacal silver staining which is incompatible with the tricine system. Second, the system offers a great versatility with minimal changes in pH and %T changes. While the interest of this feature can be marginal in standard SDS-PAGE, it becomes much more interesting in large scale 2D electrophoresis-based studies, where it is always risky and cumbersome to introduce major changes in separation protocols. For example, gels intended to resolve low molecular weight proteins and gels intended to resolve high molecular weight proteins can be run in a single multi-plate tank with only a minor difference in running time. Furthermore, the use of blank plates can allow optimal running of the slower gels once the faster ones have been removed.

We thus investigated the performances of the taurine system in 2D electrophoresis. Figure 2 shows typical results obtained in 10%T, 2.7%C gels with the taurine trailing ion at various pH. Further modulation of the Mw range analyzed on the gels can be obtained by coupling appropriate pH values with minimal modulation of the acrylamide concentration. Figure 3 shows the effect of lowering the %T from 10 to 9% on the resolution of high Mw proteins, while figure 4 shows the effect of increasing the %T from 10 to 11.5 % on the resolution of low Mw proteins. Here again, important modulation can be obtained with small changes in %T. This allows in turn minimal changes in gel strength from one experiment to another, and minimal changes in gel porosity, thereby allowing very reproducible protease digestions from one gel to another for peptide mass fingerprinting experiments. This allowed in turn efficient protein characterization over the whole Mw range, as shown in figure 5. As a test, we investigated the sequence coverage obtained on low molecular weight proteins to the one obtained in the tricine or glycine systems with 15%T gels. The results are shown on table 1, and demonstrate that the sequence coverage is always at least as good with the taurine system that the ones obtained with the tricine and glycine systems.

This remarkable versatility of the taurine system in both directions (i.e. toward the high molecular weight proteins and toward the low molecular weight proteins), is not found in the glycine and tricine systems. We attribute it to the fact that there is a good adequation between the pK of Tris and the pH at which the taurine mobility matches the mobility of proteins, which is not the case with glycine and tricine. In the case of tricine, the relatively low pK of its amino group (8.4) makes it very mobile at a lower pH than taurine. While this is a great advantage for the analysis of low molecular weight proteins [7], this is clearly a problem for the analysis of high molecular weight proteins, where a pH of 7 or below is required [10]. This is hardly reached with a Tris-based buffer, which further complicates the analysis and increases its cost. The converse situation is observed with glycine, with the high pK of its amino group (9.7). This requires a high pH buffer (8.8 in the Ornstein system), and the further increase in pH required for the analysis of low molecular weight proteins goes well beyond the buffering power of Tris [5]. However, reduction of the pH is easy in this case and allows good resolution of high molecular weight proteins [4, 19]

Concluding remarks

The electrophoresis system introduced here meets the requisites we had fixed to it. It allows to analyze easily proteins from 5 to 250 kDa with a single electrode buffer, various gels pH and minimally varying acrylamide concentrations. This allows even performances in electrophoresis and in post-electrophoretic processes such as staining and identification by peptide mass fingerprinting. In addition, this versatility allows to run in parallel with a single cathode buffer gels intended to separate proteins of varied size. Moreover, this gel system is compatible with all the staining procedures designed for standard SDS gels, which normally use the glycine system.

Acknowledgements
TR wants to acknowledge personal support from the CNRS

Legends to figures

Figure 1: Analysis of molecular weight markers
Molecular weight marker proteins were analysed by SDS-PAGE in various electrophoretic systems. The markers are: on the left lane, Bio-Rad wide range markers, in the middle lane, albumin + Myoglobin-CNBr peptides, and in the right lane, when present, albumin crosslinked with pyromellitic dianhydride. The molecular weight is indicated close to the corresponding band. Proteins were detected by silver staining (non-ammoniacal).
Panel A: Taurine trailing ion, 7.5 %T gel, Tris 110/0.6; Panel B: Taurine trailing ion, 9 %T gel, Tris 110/0.6; Panel C: Taurine trailing ion, 10 %T gel, Tris 110/0.6; Panel D: Glycine trailing ion, 10 %T gel, Tris 240/0.3 (i.e. pH 8.8); Panel E: Tricine trailing ion, 10 %T gel, Tris 240/0.6 (i.e. pH 8.4); Panel F: Taurine trailing ion, 10 %T gel, Tris 130/0.6; Panel G: Glycine trailing ion, 15 %T gel, Tris 240/0.3 ; Panel H: Tricine trailing ion, 10 %T gel, Tris 240/0.6; Panel I: Taurine trailing ion, 10 %T gel, Tris 150/0.6; Panel I: Taurine trailing ion, 11.5 %T gel, Tris 150/0.6;

Figure 2: 2D electrophoresis of whole cell extracts (acidic and neutral proteins)
Whole cell extracts prepared from HeLa cells were analysed by 2D electrophoresis. The pH gradient in the first dimension ranged from 4 to 8, thereby separating the acidic and neutral cellular proteins only. The gel buffer varied from Tris 110/0.6 (theoretical pH 7.75, panel A) to Tris 130/0.6 (theoretical pH 7.95, panel B and to Tris 150/0.6 (theoretical pH 8.05, panel C). Equivalent proteins in the 3 gels are pointed by arrows.

Figure 3: Analysis of high molecular weight proteins
Whole cell extracts prepared from HeLa cells were analysed by 2D electrophoresis. Tris 110/0.6 was used in both gels. Panel A: 10%T gel. Panel B: 9%T gel

Figure 4: Analysis of low molecular weight proteins
Whole cell extracts prepared from HeLa cells were analysed by 2D electrophoresis. Tris 150/0.6 was used in both gels. Panel A: 10%T gel. Panel B: 11.5%T gel

Figure 5: Protein identification by peptide mass fingerprinting.
Proteins in various areas of the gels were excised and submitted to in-gel protease digestion and subsequent analysis of the extracted peptides by mass spectrometry.

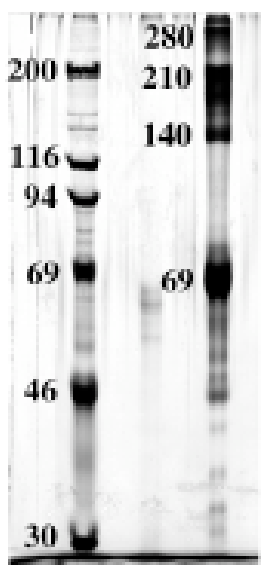
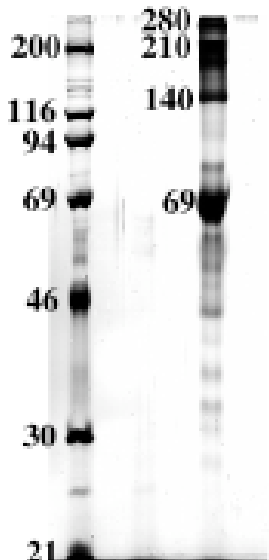
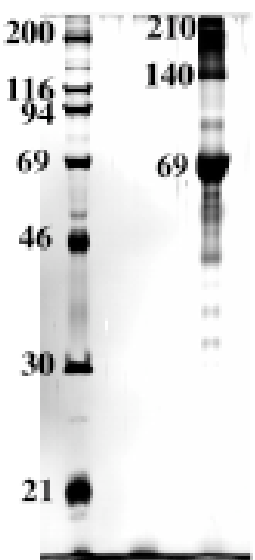
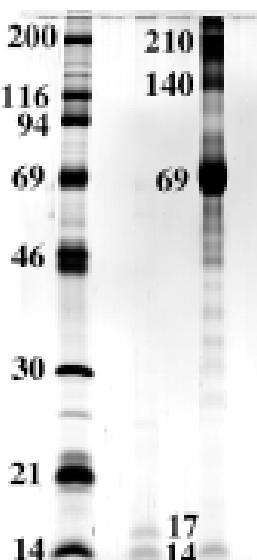
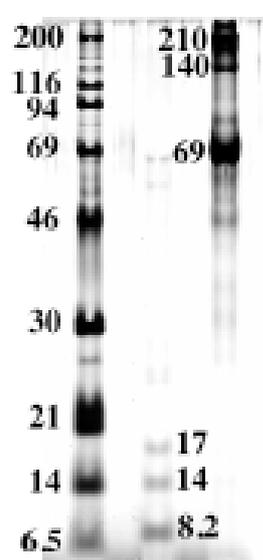
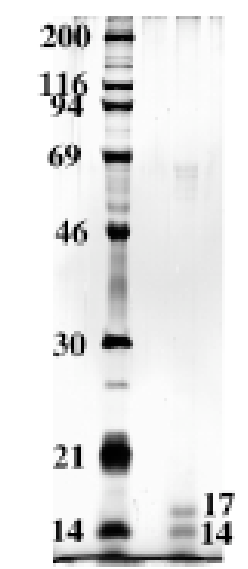
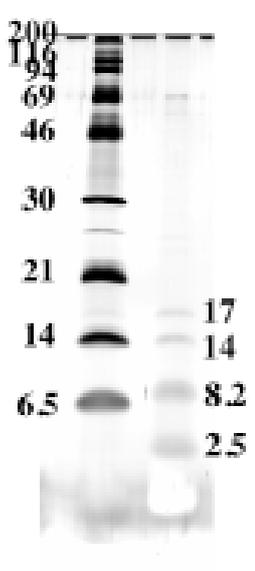
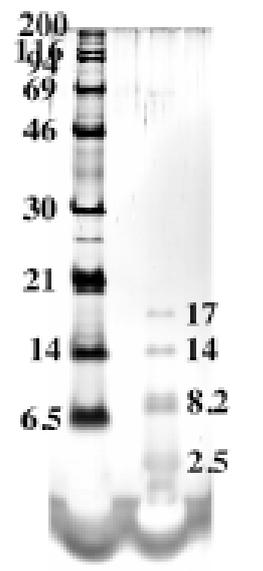
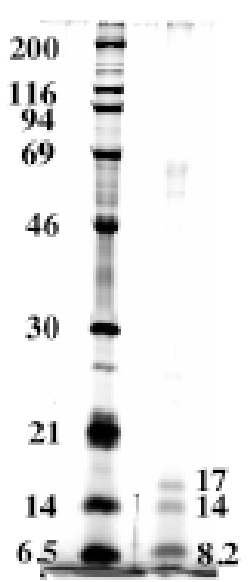
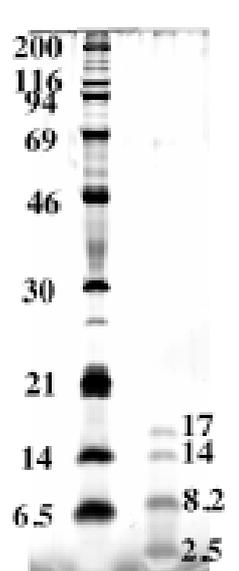

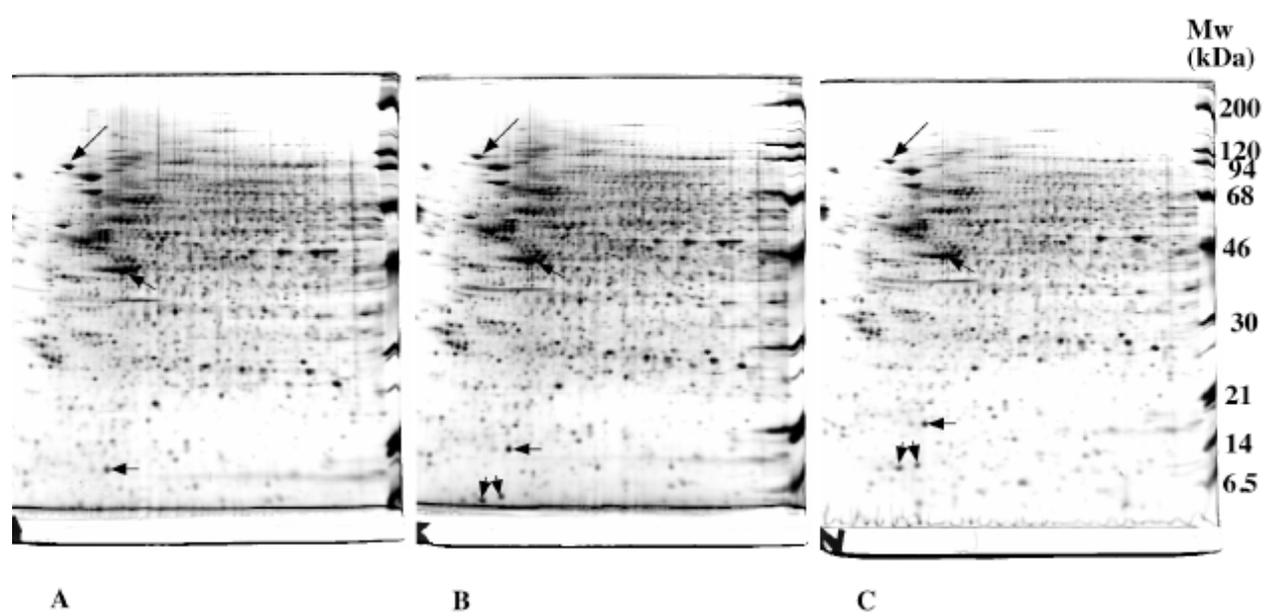

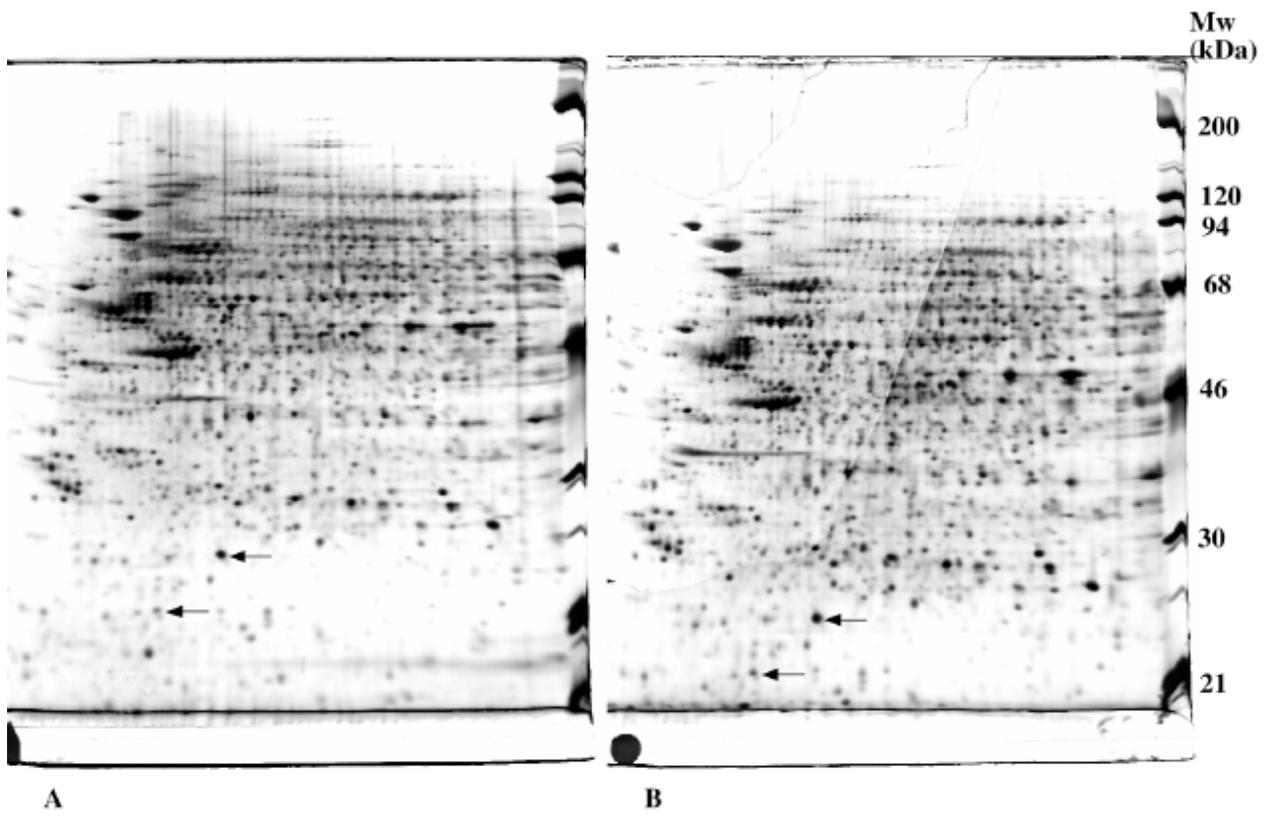

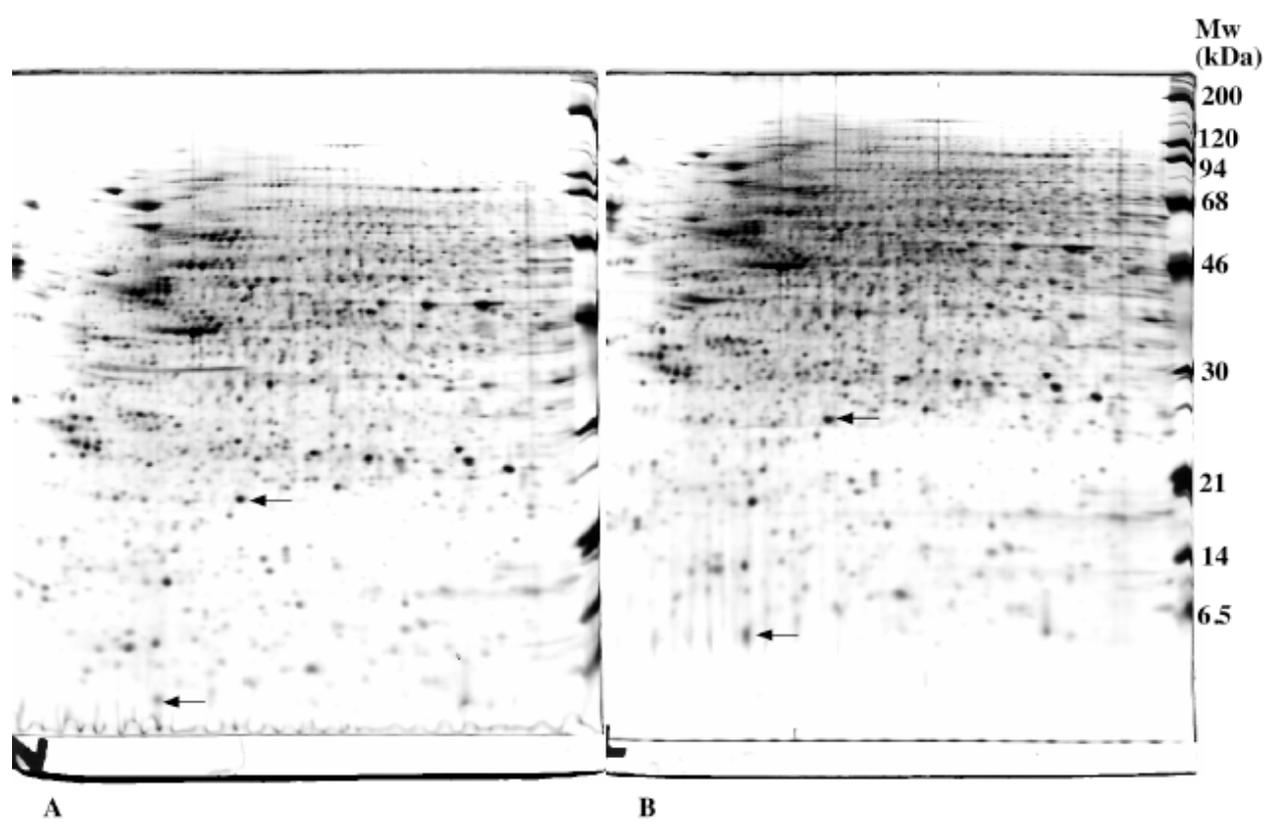

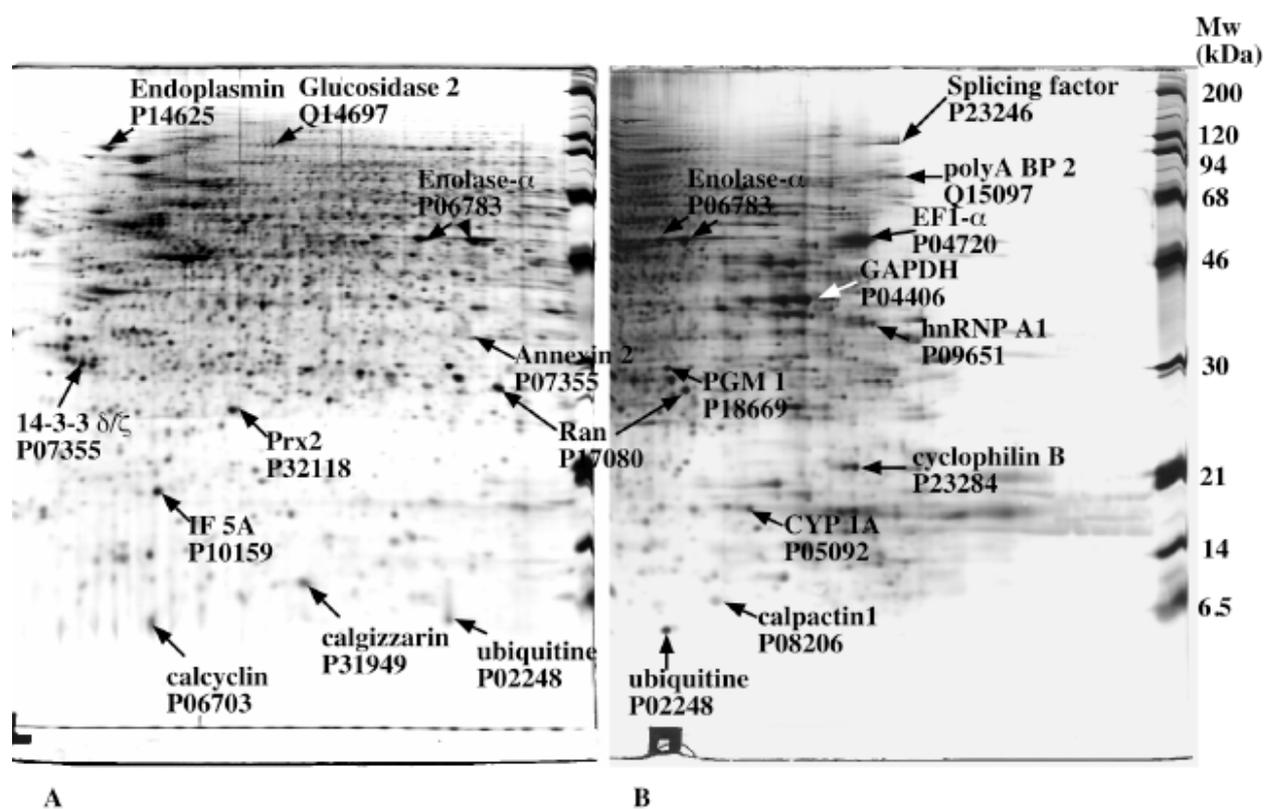

Table 1: protein identification by mass spectrometry

| Protein name | Accession number | Separation system | Theoretical Mw/pI | coverage |
|---|---|---|---|---|
| GTP Binding protein RAN | P17080 | taurine | 24579 / 7.01 | 63% |
| Peroxiredoxin 2 | P32118 | taurine | 22049 / 5.66 | 40% |
| Initiation factor 5A | P10159 | taurine | 16918 / 5.08 | 34% |
| Annexin II | P07355 | taurine | 38677 / 7.56 | 55% |
| 14-3-3 Protein zeta/delta | P29312 | taurine | 27899 / 4.73 | 41% |
| Endoplasmin | P14625 | taurine | 92697 / 4.76 | 28% |
| Glucosidase II | Q14697 | taurine | 107289 / 5.71 | 25% |
| Peptidyl-prolyl cis-trans isomerase B | P23284 | taurine | 22785 / 9.33 | 46% |
| PolyA binding protein2 | Q15097 | taurine | 58709 / 9.31* | 41% |
| Splicing factor | P23246 | taurine | 76216 / 9.45 | 24% |
| Enolase alpha | P06733 | taurine | 47350 / 6.99 | 61% |
| phosphoglycerate mutase | P18669 | taurine | 28769 / 6.75 | 68% |
| GAPDH | P04406 | taurine | 36070 / 8.58 | 51% |
| hn RNP A1 | P09651 | taurine | 38805 / 9.26 | 41% |
| Peptidyl-prolyl cis-trans isomerase A | P05092 | taurine | 18098 / 7.82 | 39% |
| calpactin 1 | P08206 | taurine | 11064 / 7.30 | 52% |
| calpactin 1 | P08206 | glycine | 11064 / 7.30 | 28% |
| calpactin 1 | P08206 | tricine | 11064 / 7.30 | 34% |
| ubiquitin | P02248 | taurine | 8560 / 6.56 | 61% |
| ubiquitin | P02248 | glycine | 8560 / 6.56 | 50% |
| ubiquitin | P02248 | tricine | 8560 / 6.56 | 61% |
| calgizzarin | P31949 | taurine | 11733 / 6.56 | 56% |
| calgizzarin | P31949 | glycine | 11733 / 6.56 | 52% |
| calgizzarin | P31949 | tricine | 11733 / 6.56 | 54% |
| calcyclin | P06703 | taurine | 10173 / 5.33 | 45% |
| calcyclin | P06703 | glycine | 10173 / 5.33 | 26% |
| calcyclin | P06703 | tricine | 10173 / 5.33 | 26% |

*: the protein deposited in the database is the N-terminal fragment of the complete protein